\documentclass[10pt, letterpaper]{IEEEtran}
\usepackage{amsmath}
\usepackage{graphicx,subfigure,comment}
\usepackage[table,dvipsnames]{xcolor}
\usepackage{url} 
\usepackage{dsfont}
\usepackage{cite}
\usepackage{tabularx}
\usepackage{dblfloatfix}

\newcommand{\Fig}[1]{Fig.~\ref{fig:#1}}

\newcommand{\Sec}[1]{Sec.~\ref{sec:#1}}

\begin{document}

\title{Towards Failure Resiliency in 5G: Service Shifting}

\author{
    \IEEEauthorblockN{Francesco Malandrino\IEEEauthorrefmark{1}\IEEEauthorrefmark{2}\IEEEauthorrefmark{3}, Carla-Fabiana Chiasserini\IEEEauthorrefmark{2}\IEEEauthorrefmark{1}\IEEEauthorrefmark{3}, Giada Landi\IEEEauthorrefmark{4}}
    \\
    \IEEEauthorblockA{\IEEEauthorrefmark{1}CNR-IEIIT, Torino, Italy}
    \IEEEauthorblockA{\IEEEauthorrefmark{2}DET, Politecnico di Torino, Italy}
    \IEEEauthorblockA{\IEEEauthorrefmark{3}CNIT, Parma, Italy}
    \IEEEauthorblockA{\IEEEauthorrefmark{4}NextWorks s.r.l., Pisa, Italy
    \\francesco.malandrino@ieiit.cnr.it, chiasserini@polito.it, g.landi@nextworks.it}
}

\maketitle

\pagestyle{plain}

\begin{abstract}
Many real-world services can be provided through multiple VNF graphs, corresponding, e.g., to high- and low-quality variants of the service itself. Based on this observation, we extend the concept of service scaling in network orchestration to {\em service shifting}, i.e., switching between the VNF graphs implementing the same service. Service shifting can serve multiple goals, from reducing operational costs to reacting to infrastructure problems. Furthermore, it enhances the flexibility of service-level agreements between network operators and third party content providers (``verticals''). In this paper, we introduce and describe the service shifting concept, its benefits, and the associated challenges, with special reference to how service shifting can be integrated within real-world 5G architectures and implementations. We conclude that existing network orchestration frameworks can be easily extended to support service shifting, and its adoption has the potential to make 5G network slices easier for the operators to manage under high-load conditions, while still meeting the verticals' requirements.
\end{abstract}

\section{Introduction}
\label{sec:intro}

5G networks are built {\em for services}, not merely for connectivity. Third-party providers, called {\em verticals} (e.g., automotive industries, e-health companies, and media content providers), will purchase from mobile operators the networking and processing capabilities necessary to provide their services. Such services will concurrently run on the mobile operator's infrastructure, which will support their diverse requirements under the so-called {\em network slicing} paradigm~\cite{slicing1,slicing2}.

Services are specified by verticals~\cite{slicing1,slicing2} as a set of virtual network functions (VNFs) connected to form a VNF forwarding graph (VNFFG), along with the needed target  Key Performance Indicators (KPIs), e.g., maximum delay or minimum reliability. Operators will host the VNFs on their own infrastructure, ensuring that they are assigned enough resources for the service to meet the target KPIs while keeping operator costs as low as possible. Such a problem is known as network orchestration~\cite{slicingalgos} or VNF placement~\cite{noi-infocom18}.

It is a natural and often unspoken assumption that every vertical service is associated with {\em one} VNF graph\footnote{Possibly including subgraphs corresponding to nested services~\cite{slicing2}.}: either the service can be provided through the specified VNFs with the target KPIs, or the service deployment fails. In some cases, resource shortages are managed by limiting the damage, e.g., getting as close as possible to the target KPIs~\cite{noi-infocom18} or enforcing different priorities among services; however, it is typically assumed that VNFs composing a service requested by a vertical are not changed.

On the contrary, in many real-world cases, such as those discussed in \Sec{relevance}, the same vertical service can be provided  through a full-fledged, {\em primary} VNF graph, and also in a suboptimal yet useful fashion through a different, {\em secondary} graph. The mobile operator can thus perform two additional operations when matching the services to provide with the available resources: it can {\em shift down} a certain service, dropping its primary VNF graph and deploying the secondary one in case of resource shortage, or {\em shift up} that service performing the opposite operation.

In this paper, we discuss the role of service shifting operations in 5G networks, as well as the opportunities and challenges they bring. Specifically, \Sec{relevance} discusses the relevance of shifting operations, presenting several examples of services that can benefit from them, while \Sec{relwork} reviews the main related works. \Sec{opportunities} and \Sec{challenges} discuss the role of service shifting decisions in a comprehensive network orchestration strategy and the associated challenges. Finally, \Sec{conclusion} concludes the paper and sketches future research directions.

\section{Shifting services}
\label{sec:relevance}

Generally speaking, shifting vertical services up or down is possible when the same goal can be pursued through different strategies, associated with significantly different resource requirements.

\begin{figure}
\centering
\subfigure[\label{fig:vnffg-sensors}]{
\includegraphics[width=1\columnwidth]{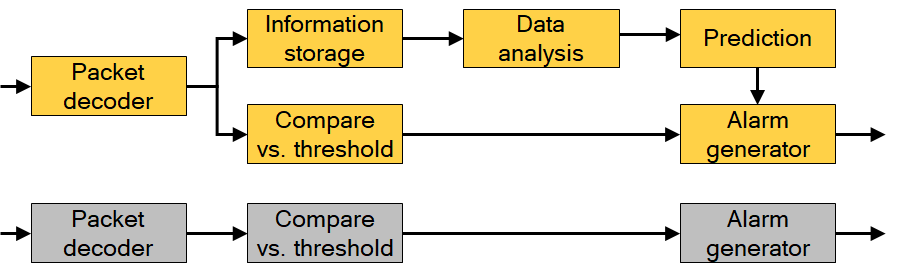}
} 
\subfigure[\label{fig:vnffg-seethrough}]{
\includegraphics[width=1\columnwidth]{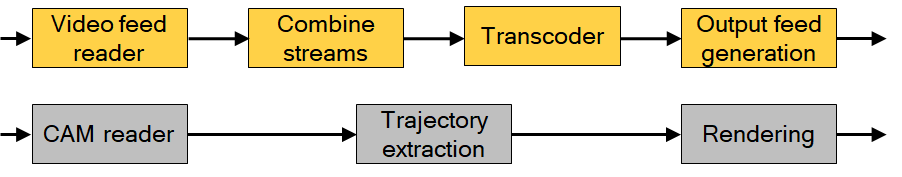}
} 
\caption{
Primary (gold background) and secondary (silver background) VNF graphs associated with a sensor monitoring (a) and a vehicular see-through (b) service. Notice that some VNFs may be common to both graphs, as in (a).
\label{fig:vnffg}
} 
\end{figure}

A good example is the sensor monitoring service presented in \Fig{vnffg-sensors}: in ordinary conditions, sensor readings are checked against static thresholds and used for prediction. An alarm is generated if current values exceeded the static threshold or the predicted values are detected as anomalous. However, if a resource shortage prevents the primary VNF graph from being deployed, there is a benefit in {\em at least} being able to raise an alarm if thresholds are exceeded, by implementing the bottom VNF graph in \Fig{vnffg-sensors}. Implementing such a {\em secondary} graph is preferable, for both the vertical and the mobile operator, to not implementing the service at all.

The vehicular see-through service presented in \Fig{vnffg-seethrough} provides a second example of such a situation. Large vehicles, e.g., trucks, are equipped with cameras capturing their view of the road. Such video streams are transmitted to the infrastructure where they are read, combined, transcoded and served to the vehicles whose view would otherwise be obstructed. The VNFs required by such a service, represented in the top part of \Fig{vnffg-seethrough}, require significant computational and bandwidth resources. If such resources are not available, the see-through service can be implemented through the VNFs in the lower part of \Fig{vnffg-seethrough}. Therein, instead of relying on video streams, cooperative awareness message (CAMs) are leveraged to construct a schematic view of the positions of the preceding vehicles. The two VNF graphs serve the same purpose -- warning following vehicles of potential hazards -- in different ways, associated with different levels of resource consumption.

{\bf Generalization.}
It is worth mentioning that, in general, service shifting does not require necessarily {\em distinct} primary and secondary VNF graph. One graph can be a subset of the other, as in \Fig{vnffg-sensors}, or they can be completely disjoint as in \Fig{vnffg-seethrough}, or any intermediate combination thereof, with only some VNFs being in common between the two graphs. 
Furthermore, for sake of simplicity, in this paper we only refer to services with one primary and one secondary VNF graph; however, such a notion can be easily extended. As an example, some services may be associated with more than two graphs, e.g., primary, secondary, and (so to say) tertiary ones. Alternatively, a service may have one primary graph and two secondary ones, requiring, respectively, less networking resources and less computational resources. In the most general terms, each service may be associated with any number of VNF graphs, each with different resource requirements and a different degree of usefulness for the vertical.

\section{Related work}
\label{sec:relwork}

Our work is related to three main research areas: network slicing architectures and applications, network orchestration and VNF placement algorithms, and reliability-aware orchestration.

A first group of works, including~\cite{slicing1,slicing2}, focus on establishing a link between the new use cases for 5G network, their requirements (e.g., the need to concurrently support multiple vertical services), and network slicing. Specifically, \cite{slicing1}~focuses on cloud Radio Access Network (RAN) scenarios, and remarks how network slicing is able to simplify the management of user mobility across access networks and the associated resource allocation decisions. The authors of~\cite{slicing2} take the viewpoint of a network operator, and discuss how network slicing can simplify the creation of multiple, virtualized access networks with different speed, latency, and reliability requirements. Taking the same viewpoint, \cite{contreras2017network}~compares the main options for the management of network slices, i.e., provider-managed and tenant-managed slices.

A substantial body of works is dedicated to the decisions required in network slicing scenarios, i.e., the network orchestration problem. The work in~\cite{slicingalgos} identifies the unique algorithmic challenges associated with network slicing, including the need to account for different types of constraints -- from end-to-end delays to multi-tenancy and isolation issues. Furthermore, it presents a low-complexity solution concept for real-time network slicing, based on monitoring and forecasting the state of the network, and on an efficient, two-phase, online optimization. The study in \cite{bagaa2018coalitional} focuses on a core network as a service (CNaaS) scenario, where multiple verticals share their virtual EPC (vEPC) instances. The high-level objective is to satisfy all the verticals' requirements with the smallest possible number of vEPC instances, hence, the lowest cost for the operator. To this end, the authors resort to cooperative game theory, and study how to build coalitions of verticals sharing the same vEPC instance.

Many works, including~\cite{draxler2018jasper,noi-infocom18} seek to {\em jointly} make the decisions required for network orchestration, i.e., VNF placement, VNF resource assignment, and traffic routing. In both \cite{draxler2018jasper,noi-infocom18}, the rationale is that such decisions impact each other, and it is thus necessary to account for their interaction. The two works have different underlying assumptions (as an example, the CPU assigned to each VNF is static in~\cite{draxler2018jasper} and dynamic in~\cite{noi-infocom18}) and use different methodologies (namely, graph theory in~\cite{draxler2018jasper} and queuing theory in~\cite{noi-infocom18}). Finally, several works propose algorithmic approaches tailored to a specific application of network slicing: examples include 
\cite{dao2018sgco}, which focuses on Internet-of-things (IoT) scenarios and seeks to make energy-efficient orchestration decisions.

Especially relevant to our study are those works that take into account reliability and survivability in 5G networks. Among these, \cite{petrov2018achieving} focuses on a vehicular scenario where multiple access networks are available, e.g., mmWave and Wi-Fi. In such a context, the reliability of individual wireless links is estimated, and mission-critical traffic is routed through the link {\em or links} whose aggregate reliability matches the requirements. \cite{qu2018reliability} studies how to combine unreliable individual VNFs into a reliable service chain. The basic approach is to enhance reliability through duplication, e.g., deploying two instances of the same VNF so that if one fails the other can take over. However, this can lead to unused resources and higher-than-necessary cost. To counter this, the authors formulate an optimization problem yielding the minimum-cost duplication decisions consistent with reliability targets. \cite{shahriar2018virtual} takes an opposite approach to a similar problem, and aims at augmenting the VNF graph, e.g., by duplicating some parts thereof, to obtain the required reliability level.

With the exception of~\cite{shahriar2018virtual}, {\em all} the above works assume that VNF graphs are given and immutable; furthermore, \cite{shahriar2018virtual}~itself envisions to perform some operations on the one VNF graph given as an input, as opposed to having multiple VNF graphs providing the same service. It is also worth remarking that our service shifting approach can be used to pursue any goal, be it cost minimization (as in~\cite{bagaa2018coalitional,draxler2018jasper,zhang2018scalable,dao2018sgco}), reliability/survivability (as in~\cite{petrov2018achieving,qu2018reliability,shahriar2018virtual}), or a combination of the two.

\section{Applications and decision-making approaches}
\label{sec:opportunities}

This section describes two of the main applications of service shifting, namely, reacting to resource shortage situations (\Sec{sub-shortage}) and extending the expressiveness of Service Level Agreements (SLAs) (\Sec{sub-cost}). For each application we also discuss the decision-making entities that are involved and the approaches they can take.

\subsection{Reaction to resource shortage}
\label{sec:sub-shortage}

As mentioned earlier, in 5G networks operator-owned resources (e.g., servers) are used to run vertical-specified services, i.e., the VNFs composing their VNF graph. A {\em resource shortage} situation happens when the quantity of available resources drops unexpectedly, or the traffic load grows suddenly. This can be caused by several different conditions, including:
\begin{itemize}
    \item problems in the operator infrastructure, e.g., servers breaking down or data centers becoming inaccessible due to link failures;
    \item sudden increases in traffic, including mass events (``flash crowds'');
    \item emergency situations and natural disasters, whereby parts the network infrastructure can be destroyed and network demand, by both victims and responders, increases.
\end{itemize}

In resource shortage conditions, the operator is unable to meet all target KPIs for all services. The traditional approach is to {\em re-orchestrate}~\cite{shortage-wcncw} the affected services, which include (i) moving VNFs from unavailable servers to operating ones, and (ii) scaling down the resources they are assigned. This unavoidably results in KPI targets being violated, which, in turn, may jeopardize the usefulness of the service itself, e.g., lagging video for the see-through service discussed in \Sec{relevance}.

In this context, service shifting represents a very attractive alternative to scaling down. Instead of trying to implement the primary VNF graph of a service while missing the associated KPI targets, the operator can shift down that service and provide it through its secondary VNF graph. As for choosing {\em which} services to shift down, the operator can follow several approaches, including:
\begin{itemize}
    \item payoff maximization: down-shifted services bring a reduced revenue, hence, shift down the services associated with the lowest revenue loss;
\item minimization of the user QoE degradation: down-shifted services result in  a lower user satisfaction as the quality of experience users perceive may be severely impacted, hence shift down the less popular services;
    \item  minimization of the service reaction time: re-orchestration, e.g., instantiating new VNF instances and updating routing tables, takes a non-negligible time, hence, shift down the services requiring the fewest such operations.
\end{itemize}

\subsection{Extending SLAs}
\label{sec:sub-cost}

The possibility of service shifting can be leveraged during the SLA negotiation between verticals and operators. As an example, a vertical may accept that the secondary VNF graph is used for its service for a certain fraction of requests and/or in certain times of the day, in exchange of a reduced fee. Similarly, the semantics of service priorities can be extended to mandate that a service can be shifted down only if all lower-priority services (by the same vertical) have already been shifted down.

For operators, service shifting means extending the orchestration options: in addition to VNF placement and resource assignment~\cite{noi-infocom18}, operators will be able to use shifting decisions to pursue their high-level objective to meet the  SLA commitments while minimizing costs. For verticals, service shifting is an additional way to express their needs when negotiating SLAs, thus avoiding paying for unnecessary resources or features. 

On the negative side, orchestration decisions are bound to become more complex, from several viewpoints. One is the sheer computational complexity, the other number of orchestration options. This, in turn, can make additional decision-making entities necessary, which make the network architecture more complex. Finally, such entities need large quantities of up-to-date information to make high-quality decisions, and collecting and processing such information may be a non-trivial task. We discuss all such challenges in \Sec{challenges}, along with possible solutions.

\section{Challenges}
\label{sec:challenges}

This section discusses three of the main challenges associated with service shifting, including: identifying the appropriate decision-making entities (\Sec{sub-decisions}), gathering the necessary information through service and network monitoring (\Sec{sub-monitoring}), and swiftly deploying or re-deploying the shifted services (\Sec{sub-redeployment}).

\subsection{Decision-making entities and interfaces}
\label{sec:sub-decisions}

Service shifting can be viewed as an extension to traditional network orchestration, which makes orchestration decisions even more complex to handle. This further strengthens the need, recently emerged in the 5G research community, to distribute the burden of network orchestration decisions across multiple decision-making entities, working at different abstraction layers.

\begin{figure}
\centering
\includegraphics[width=1\columnwidth]{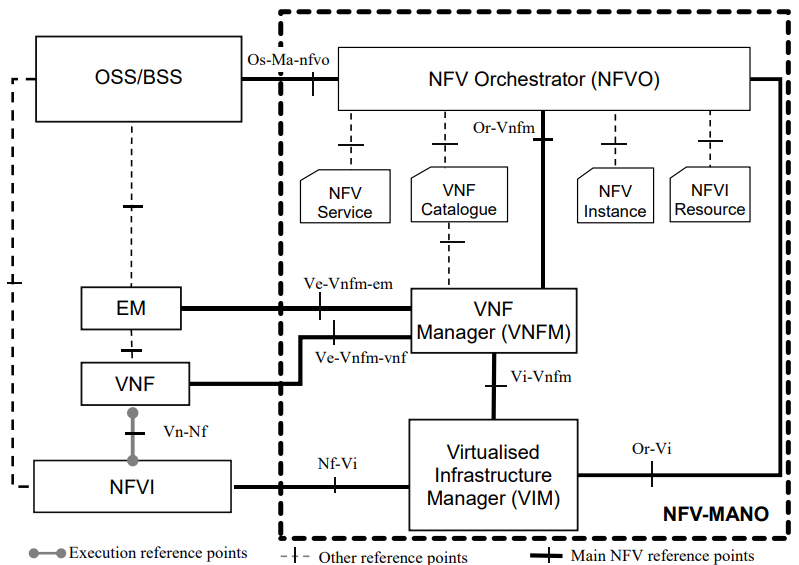}
\caption{
    The NFV-MANO architectural framework as standardized by ETSI. Source: ETSI GS NFV MAN 001.
    \label{fig:mano}
} 
\end{figure}
\begin{figure*}
\centering
\includegraphics[width=1.7\columnwidth]{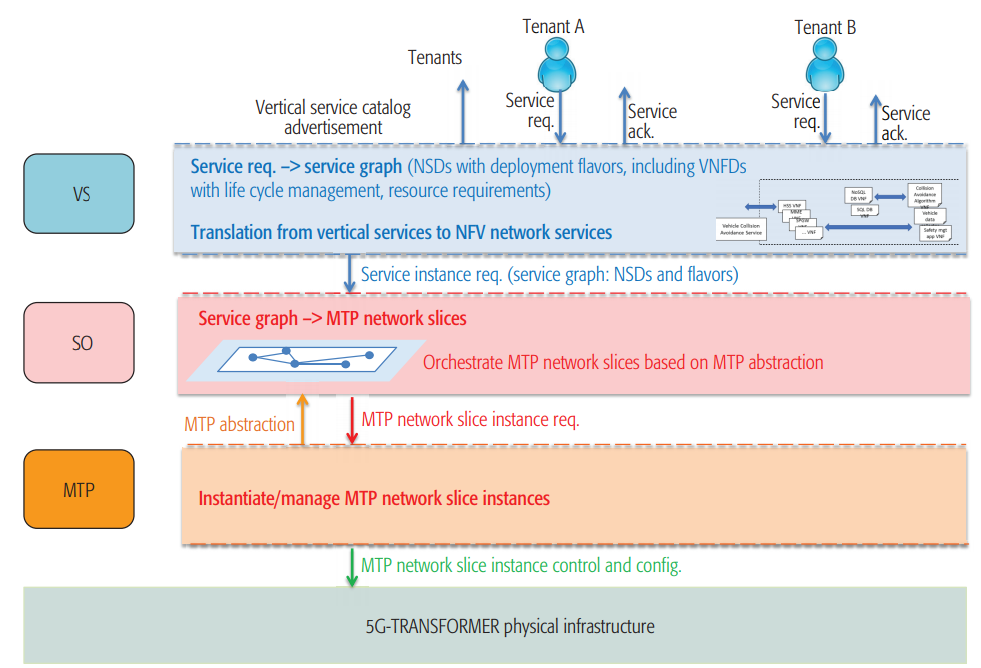}
\caption{
    The high-level architecture and interactions proposed by the 5G-TRANSFORMER project. Source:~\cite{5gt}.
    \label{fig:transformer}
} 
\end{figure*}

In the network management and orchestration (MANO) framework, standardized by ETSI in standard GS NFV MANO 001 and represented in \Fig{mano}, virtually all network orchestration decisions are made by the NFV Orchestrator (NFVO). The NFVO takes as an input the service graphs and KPIs specified by verticals through the Operation and Business Support Services (OSS/BSS); its output is represented by VNF instantiation and placement decisions, which are subsequently enacted by lower-level entities like the VNF manager (VNFM).

Several 5G-related research efforts envision alternative solutions, advocating to split the tasks assigned to the NFVO in \Fig{mano} between two entities: a higher-level one, making decisions on a per-service basis, and a lower-level one, working with individual VNFs with decisions more oriented to resource-based criteria. Taking the architecture proposed by the H2020 project 5G-TRANSFORMER in~\cite{5gt}, and represented in \Fig{transformer}, we can identify:
\begin{itemize}
    \item a vertical slicer (VS), translating the verticals' requirements into service graphs, also accounting for the service-level agreements (SLAs) in place;
    \item a service orchestrator (SO), taking the service graph as an input and using the network, computing and storage resources available in the infrastructure to build the network slice that will run the service.
\end{itemize}
In such a context, service shifting decisions can be made by higher-level, service-aware entities such as the VS. This avoids further increasing the burden on lower-level entities like the SO, which are already in charge of VNF placement and resource assignment.

Both high- and low-level entities have challenging decisions to make. High-level entities must ensure that the SLAs with verticals are honored, which also requires them to check whether shifting a certain service is permitted under its SLA. In case no shifting is permitted for any service but some shifting is needed, e.g., in emergency conditions, the high-level entity must choose the SLA to violate, -- possibly the one associated with the lowest monetary or safety penalty. As far as low-level entities are concerned, they are faced with a higher request volume, i.e., a lager number of decisions to make, due to the switching between primary and secondary graphs of the same service. Furthermore, such decisions are often more complex than those related to ordinary service scaling, since they may involve deploying a completely new VNF graph, as opposed to making minor adjustments to currently-deployed ones, e.g., deployment flavor changes.

Having two decision-making entities instead of one unavoidably brings additional issues, connecting with the need to coordinate them. This, in turn, requires (i) gathering and sharing the information they need to make their decisions, and (ii) swiftly propagating such decisions to the entities in charge of enacting them.

This has an impact on the interfaces between the components of the extended MANO framework and with the monitoring platform adopted to collect data from the different sources. In particular, suitable policies should be exchanged and configured across the different MANO modules to regulate the kind of shifting decisions that should be applied to a given service, as well as the designated decision point. Network service and slice descriptors should also encode the criteria to be adopted for such decisions, identifying the monitoring parameters to be collected and evaluated, as well as the shifting rules with triggering conditions and target actions. Moreover, suitable primitives must be defined between VS and SO to handle the delegation of shifting decisions between the two entities and to maintain their synchronization about the current service status and its resource usage. This can be achieved through mechanisms for asynchronous notifications about shifting decisions and actions taken on the lifecycle, the functional components, and the virtual resources of the service in each of the MANO layers. 

The interface with the monitoring platform, on the other hand, should allow both VS and SO to collect a set of configurable monitoring information, even in the form of notification alerts. The target monitoring data, as further elaborated in the following section, cover different scopes to provide input to entities that operate at different levels: measurements related to infrastructure and resource usage will feed SO algorithms, while application- and service-based monitoring information will be used at the VS level.

The remainder of this section is devoted to analyze in detail the issues related to the collection and sharing of monitoring data and to the enforcement and propagation of service shifting decisions, highlighting how these functionalities can be tackled within current 5G architectures.

\subsection{Monitoring}
\label{sec:sub-monitoring}

\begin{figure}
\centering
\includegraphics[width=1\columnwidth]{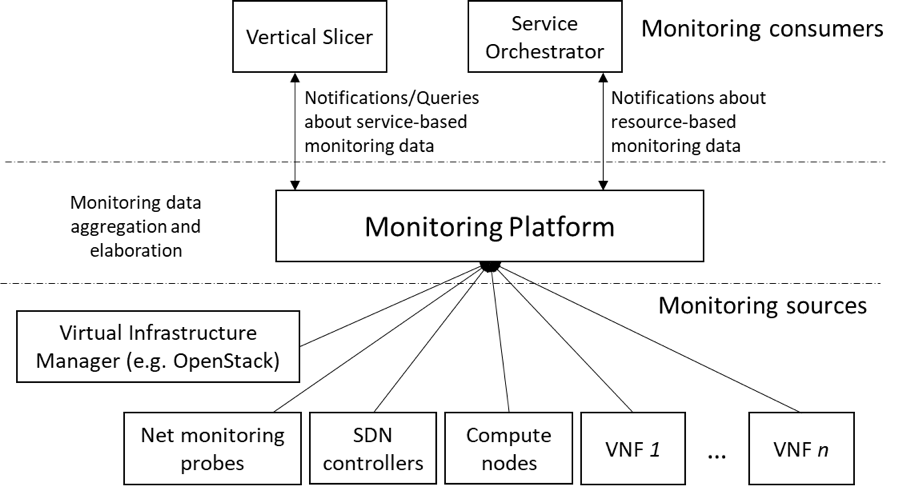}
\caption{
    Monitoring platform as collector of data from several monitoring sources.
    \label{fig:monitor}
} 
\end{figure}

As mentioned in \Sec{sub-decisions}, the decision-making entities at the VS and SO run algorithms that need to receive as input different kinds of monitoring data, related to a variety of physical and virtual components and resources, from physical infrastructures to virtual resources, up to application and service level data. The monitoring platform should be flexible enough to support different types of customizable data sources in a distributed environment, as shown in \Fig{monitor}. They should also  implement preliminary data elaboration tasks to efficiently deliver aggregated monitoring parameters and produce automated notifications, based on simple thresholds or more complex strategies for anomaly detection.

Since different kinds of services typically require application-specific criteria to detect resource shortage conditions or critical failures, the relevant monitoring parameters should be properly encoded in their network service descriptors. This approach allows the Service Orchestrator to deploy and configure the required data sources during the service instantiation phase. For example, it may deploy dedicated network probes in the NFV infrastructure, install monitoring data agents (or exporters) in specific VNFs as part of their configuration, and activate monitoring jobs on the monitoring platform to periodically collect reporting data on performance or resource usage. Similarly, it may also configure the monitoring platform to receive alerts when specific patterns are identified in the real-time flows of raw or aggregated monitoring data.
Depending on the placement of the decision logic, i.e., at the SO or at the VS, each or both the two architectural entities may act as consumer of high-level monitoring reports or alerts.

The complexity of aggregation and elaboration of the raw monitoring data, as collected by the elementary monitoring sources, is centralized at the monitoring platform. Such processing is driven by the rules that are dynamically configured according to the network service specification, in order to detect the particular conditions triggering scaling or shifting actions. Whenever a target pattern is detected in the aggregated monitoring data, automated alerts are notified to the monitoring consumers (VS or SO) that have an active subscription for the given pattern. Notifications may be managed either through explicit messages addressed to the target entities or through a message bus approach. Starting from the received alerts, the VS or the SO will make a decision about the need of a service shifting and will trigger the required actions, as further described in section \Sec{sub-redeployment}

It should be noted that the target sources of monitoring data are usually a mix of different kinds of specialized monitoring agents that are able to retrieve measurements and statistics related heterogeneous virtual or physical resources. Some examples of target monitoring sources for network services deployed in an NFV infrastructure are represented in \Fig{monitor}. Network-related data can be collected from SDN controllers, e.g., for statistics about traffic loads in network links or packets drop on switches or particular ports. Where needed, active monitoring probes can be deployed as part of the end-to-end network service, in particular geographical locations or along the logical VNF Forwarding Graph, to collect periodical measurements about end-to-end delay and jitter. In 5G networks, the monitoring of the radio domain is also particularly important, since the performance on the radio link has a major impact on the end-to-end delivery of the service to the mobile users. In this case, radio-related monitoring data can be collected from SDN controllers operating over the radio segments or, when available, from dedicated Multi-Access Edge Computing (MEC) services dedicated to the monitoring of radio network information (RNIS, Radio Network Information Service, as defined in ETSI GS NFV MEC 012) or mobile equipment location (as defined in ETSI GS NFV MEC 013). Monitoring data related to computing and storage resources (e.g., about consumption of RAM or vCPUs) can be retrieved from hypervisors, monitoring services already available in cloud platforms like OpenStack or node agents deployed in Virtual Machines. Finally, more high-level service-oriented monitoring data need to be produced directly by the application itself and are typically retrieved through dedicated exporters running in the VNFs.

\subsection{Swift service (re)deployment}
\label{sec:sub-redeployment}

Service shifting decisions require switching from a VNF graph to another. 
This places on decision-making entity in charge of VNF placement (e.g., the SO in \Fig{transformer}) the onus of:
\begin{enumerate}
    \item removing, potentially, all VNFs of the old graph;
    \item instantiate the VNFs required for the new graph;
    \item updating the routing rules accordingly.
\end{enumerate}

In order to enact a service shifting decision, additional changes to currently-deployed VNFs may be needed. As an example, in order to have sufficient free computational capability for the VNFs of the new graph, some other VNFs, belonging to services not being shifted, may have to be moved to a different server. If this turns out to be impossible, e.g., because there is not enough spare computational capability, a ripple effect might ensue, whereby additional service scaling or shifting decisions are made.

Shifting decisions are often made in resource shortage conditions, where KPI targets are being or may be violated. Therefore, service (re)deployment decisions must be made {\em and enacted} swiftly. The first requirement, i.e., that decisions be made quickly, is at odds with the complexity of the decisions to make, which include placing multiple VNFs throughout the network infrastructure. The second requirement, i.e., that decisions be enacted swiftly, is often overlooked but very important: indeed, real-world 5G deployments show VNF instantiation times of several tens of seconds~\cite{d52}.  
Moreover, a full operation service also needs applications completely up and running in the new VNFs; this requires additional time due to the starting procedures of the processes and the initial configuration of the applications running in Virtual Machines (VMs) or Containers. Live migration of, e.g., VMs also brings a certain degree of delay, which may impact on the services that do not need to be shifted, but just moved to different servers. A report about live migration in OpenStack Ocata\footnote{http://superuser.openstack.org/wp-content/uploads/2017/06/ha-livemigrate-whitepaper.pdf} shows average measurements from nearly 50 seconds up to 270 seconds for the time required to migrate VMs with large deployment flavors, depending on the VMs’ storage strategy (i.e., local vs. shared storage) and tunneling activation. Such delays can result in non-negligible service outage times, and substantial penalties for the mobile operator.

In order to address these issues, VNF placement algorithms shall be enhanced in two directions. First, they need to make extremely swift decisions under resource shortage conditions, even if such decisions prove to be suboptimal. Second, they need to weight the decision enactment time, e.g., VNF setup or tear-down delays, along with more traditional metrics like cost.

In summary, service shifting is compatible with the envisioned 5G architectures; however, properly exploiting its potential requires careful implementation of the decision-making entities therein and appropriate customization of the algorithms they run.

\section{Conclusion}
\label{sec:conclusion}

In this paper, we have introduced the concept of service shifting, a generalization of service scaling where the decision to make is selecting the VNF graph to use in order to provide a  service. In this context, we have discussed how and to which extent different real-world services lend themselves to shifting, the possible relationships between VNF graphs of the same service, and how service shifting can be generalized to an arbitrary number of graphs with no {\em a priori} ordering.

We have shown how service shifting can be beneficial in resource shortage situations, which may arise as a consequence of network infrastructure issues, sudden increases in traffic demand, or emergency situations. Furthermore, we have highlighted how service shifting can enhance the expressiveness and flexibility of SLAs between mobile operators and verticals, e.g., by allowing verticals to require that their service is provided through the primary VNF graph for a target fraction of  time.

Finally, we have identified the main challenges associated with service shifting, namely: (i) the additional tasks to be carried out by decision-making entities; (ii) collecting and dispatching the additional information such entities need; (iii) swiftly enacting the scaling decisions, including any needed VNF (re)deployment. Taking real-world 5G implementations as a reference, we have highlighted how such challenges can be tackled without major changes to their architecture, thus making it easy to reap the benefits of service shifting.

\bibliographystyle{IEEEtran}
\bibliography{refs}%

\section*{Acknowledgment}

This work is supported by the European Commission through the H2020 projects 5G-TRANSFORMER (Project ID 761536) and 5G-EVE (Project ID 815074).

\begin{IEEEbiography}[{\includegraphics[width=1in,height=1.25in,clip,keepaspectratio]{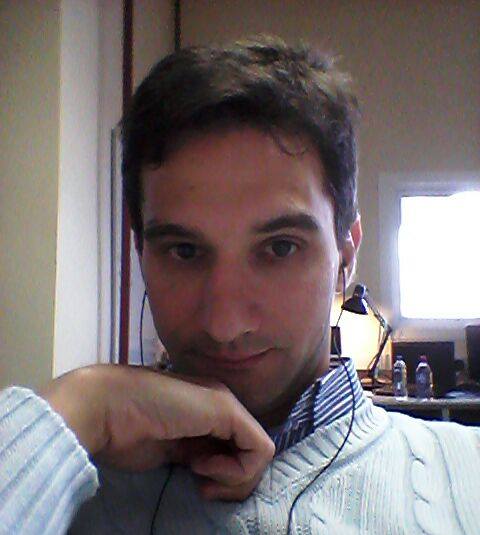}}]%
{Francesco Malandrino} received his M.S. and Ph.D. degrees from Politecnico di Torino, Italy, in 2008 and 2012 respectively. He is now a tenured researcher at the Institute of Electronics, Computer and Communication Engineering of the National Research Council of Italy (CNR-IEIIT), headquartered in Torino, Italy. Prior to his current appointment, he has been an assistant professor and a research fellow at Politecnco di Torino, a Fibonacci Fellow at the Hebrew University of Jerusalem and a research fellow at Trinity College, Dublin. His research interests include the architecture and management of wireless, cellular, and vehicular networks.
\end{IEEEbiography}

\begin{IEEEbiography}[{\includegraphics[width=1in,height=1.25in,clip,keepaspectratio]{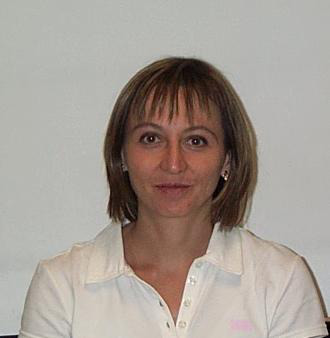}}]%
{Carla-Fabiana Chiasserini} (M'98, SM'09, F'18) graduated from
the University of Florence in 1996 and  received her Ph.D. from Politecnico di Torino, Italy, in
2000.  She worked as a visiting researcher at UCSD in 1998--2003, and as a Visiting Professor at Monash University in 2012 and 2016.  She is currently an Associate  Professor  with  the  Department  of Electronic  Engineering  and  Telecommunications  at
Politecnico di Torino.  Her research interests include architectures, protocols, and performance
analysis of wireless networks. Dr. Chiasserini has published over 300 papers in prestigious journals
and leading international conferences.
\end{IEEEbiography}

\begin{IEEEbiography}[{\includegraphics[width=1in,height=1.25in,clip,keepaspectratio]{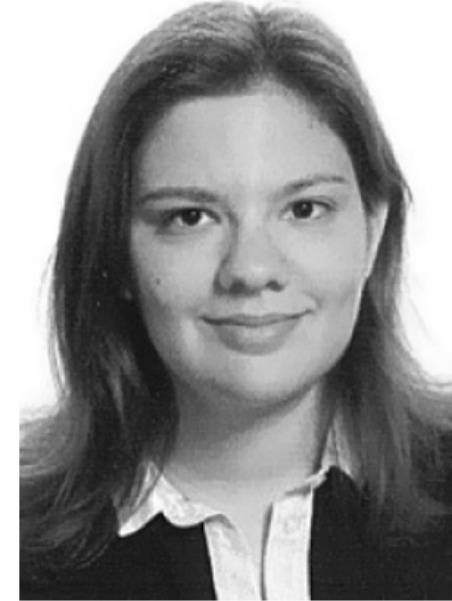}}]%
{Giada Landi} has received  the Italian  Laurea
degree ``cum laude'' in Telecommunication Engineering at the University of Pisa, Italy, in  2005.  Currently  she  is  R\&D  Project Manager at Nextworks. Her main research areas are ASON/GMPLS and PCE architectures,  Software  Defined  Networking, cloud computing, and 5G networks. Past research activities focused on SIP and IMS architecture, control plane for wireless access networks and
inter-technology mobility. She participated in  national  and  European  Projects  (FP6 WEIRD, Artemis SOFIA, FP7 GEYSERS, FP7 ETICS, FP7 CONTENT and FP7 Mobile Cloud Networking) and is currently active, among others, in the H2020 5G-TRANSFORMER and 5G-EVE projects.

\end{IEEEbiography}

\end{document}